\documentclass[aps,pra,twocolumn,showpacs, amssymb, amsmath, aps, superscriptaddress] {revtex4-2}
\usepackage{mathtools}
\usepackage{graphicx}  
\usepackage{dcolumn}  
\usepackage{bm}        
\usepackage{amsmath}
\usepackage{epstopdf}
\usepackage{latexsym}
\usepackage{mathrsfs}
\makeatletter 
\def\maketag@@@#1{\hbox{\m@th\normalfont\normalsize#1}} 
\makeatother 
\usepackage[toc,page]{appendix}  
\usepackage{graphicx}
\usepackage{amssymb}
\usepackage{amsmath}
\usepackage{algorithm} 
\usepackage[noend]{algpseudocode} 
\usepackage{amsfonts}
\usepackage[mathscr]{eucal}
\usepackage{bbold}
\usepackage{color}
\usepackage{tikz}
\usetikzlibrary{matrix,arrows,decorations.pathmorphing}

\usepackage{silence}
\newcommand{\ket}[1]{\left| #1 \right\rangle}
\newcommand{\bra}[1]{\left\langle #1 \right|}

\newcommand{\Tr}{\mathrm{Tr}}

\newcommand{\rhocoll}{\hat{\rho}_C}
\usepackage{soul}
\begin{document}

\title{Squeezing the angular momentum of an ensemble of complex multi-level atoms}

\author{D. Hemmer}\affiliation{Center for Quantum Information and Control, College of Optical Sciences and Department of Physics, University of Arizona, Tucson, AZ 85721, USA}
\author{E. Monta\~no}\affiliation{Center for Quantum Information and Control, College of Optical Sciences and Department of Physics, University of Arizona, Tucson, AZ 85721, USA}
\author{B. Q. Baragiola}\affiliation{Centre for Quantum Computation and Communication Technology, School of Science, RMIT University, Melbourne, Victoria 3001, Australia}
\affiliation{Center for Quantum Information and Control, Department of Physics and Astronomy, University of New Mexico, Albuquerque, NM 87131, USA}
\author{L. M. Norris}\altaffiliation[]{Current affiliation: Department of Physics and Astronomy, Dartmouth College, 6127 Wilder Laboratory, Hanover, NH 03755, USA.}\affiliation{Center for Quantum Information and Control, Department of Physics and Astronomy, University of New Mexico, Albuquerque, NM 87131, USA}
\author{E. Shojaee}\affiliation{Center for Quantum Information and Control, Department of Physics and Astronomy, University of New Mexico, Albuquerque, NM 87131, USA}
\author{I. H. Deutsch}\affiliation{Center for Quantum Information and Control, Department of Physics and Astronomy, University of New Mexico, Albuquerque, NM 87131, USA}
\author{P. S. Jessen}\affiliation{Center for Quantum Information and Control, College of Optical Sciences and Department of Physics, University of Arizona, Tucson, AZ 85721, USA}
\date{\today}

\begin{abstract}
\noindent Squeezing of collective atomic spins has been shown to improve the sensitivity of atomic clocks and magnetometers to levels significantly below the standard quantum limit.  In most cases the requisite atom-atom entanglement has been generated by dispersive interaction with a quantized probe field, or by state dependent collisions in a quantum gas. Such experiments typically use complex multilevel atoms like Rb or Cs, with the relevant interactions designed so atoms behave like pseudo-spin-$1/2$ particles. We demonstrate the viability of spin squeezing for collective spins composed of the physical angular momenta of $10^6$ Cs atoms, each in an internal spin-4 hyperfine state.  A peak metrological squeezing of $\gtrsim -5$dB was generated by quantum backaction from a dispersive quantum nondemolition (QND) measurement, implemented using a two-color optical probe that minimizes tensor light shifts without sacrificing measurement strength.  Other significant developments include the successful application of composite pulse techniques for accurate dynamical control of the collective spin, enabled by broadband suppression of background magnetic fields inside a state-of-the-art magnetic shield.  The absence of classical noise has allowed us to compare the observed quantum projection noise and squeezing to a theoretical model that properly accounts for both the relevant atomic physics and the spatial mode of the collective spin, finding good quantitative agreement and thereby validating its use in other contexts. Our work sets the stage for experiments on quantum feedback, deterministic squeezing, and closed-loop magnetometry. The implementation of real-time feedback may also open the door to new types of quantum simulation, wherein the evolution of a quantum system is conditioned on the outcome of a time-continuous QND measurement.  Such a scheme has the potential to access new regimes near the quantum-classical boundary, with opportunities to study long-standing issues related to quantum-classical correspondence in chaotic systems.\end{abstract}

\maketitle

\section{Introduction}
\noindent Quantum control on multiple scales, from single particles to complex many body systems, is integral to quantum information science. Examples range from digital quantum computing and analog quantum simulation to quantum metrology and sensing. For metrology and sensing applications, a rapidly growing body of work has focused on the generation and use of squeezed collective spin states to improve the performance of atomic clocks, atom interferometers, and magnetometers \cite{Pezze2018}. Recent experiments have shown that significant gains are possible in the near term, with demonstrated improvements of up to ~20 dB in the sensitivity of microwave spectroscopy relative to the standard quantum limit \cite{Hosten2016, Bohnet2014}.

Squeezing of collective atomic spins have typically been generated either through dispersive interaction with a shared mode of quantized light \cite{Hosten2016, Bohnet2014, Takano2009, Appel2009, SchleierSmith2010, Sewell2012}, or by state dependent collisions in a quantum gas \cite{Esteve2008, Riedel2010, Hamley2012, Muessel2014}. With few exceptions \cite{Takano2009, Hamley2012} these experiments have used complex atoms such as Rb or Cs, and the relevant interactions have been designed to ensure that each atom behaves, as far as possible, as an effective pseudo-spin-1/2 particle. In that case squeezing of the collective spin results solely from correlations between individual spins, and experiments can largely sidestep the complexities of quantum control within the large hyperfine manifolds that are characteristic of alkali atoms.

In this article we demonstrate that quantum backaction from a dispersive quantum nondemolition (QND) measurement can produce $\ge5$ dB of metrologically useful squeezing of a collective angular momentum formed by the physical angular momenta of individual Cs atoms in the $6S_{1/2} (f=4)$ hyperfine state. Our measurement is implemented by detecting the spin-dependent Faraday rotation of an optical probe beam during a single pass through the atomic ensemble. For hyperfine spins  $f\ge1$, this is generally far from an ideal QND scenario because of probe-induced rank-2 tensor light shifts that drive non-trivial evolution of the individual spins; this has been a major obstacle for squeezing of collective angular momenta in the past \cite{Smith2004, Deutsch2010}. Minimization of unwanted light shifts is a common challenge in atomic physics, and is usually achieved with a magic-frequency or two-component probe, for examples relevant to Faraday and QND measurements see, e. g., \cite{Chaudhury2006, Jasperse2017} and \cite{Saffman2009, Louchet-Chauvet2010}, respectively.  Here we show that the QND and nonperturbing character of the Faraday measurement can be effectively recovered through the use of a two-color probe, with components detuned relative to the Cs D1 and D2 lines in such a way that their Faraday rotation signals add constructively while the tensor light shifts cancel. Additional complications arise from the magnetic moment of individual Cs atoms, which makes initialization and squeezing of the collective angular momentum highly sensitive to ambient magnetic fields. In our experiment, we show that this problem can be addressed with a combination of magnetic shielding and composite-pulse techniques. In the end, our observed degree of squeezing is in good quantitative agreement with a model that properly accounts for the relevant atomic physics and collective spin mode, thereby validating a considerable body of theoretical work. Finally, the elimination of tensor light shifts in principle frees us to manipulate the individual hyperfine spins in ways that can enhance the entangling power of the QND measurement and maximize the overall metrological advantage \cite{Norris2012}.

\begin{figure*}
[t]\resizebox{18cm}{!}
{\includegraphics{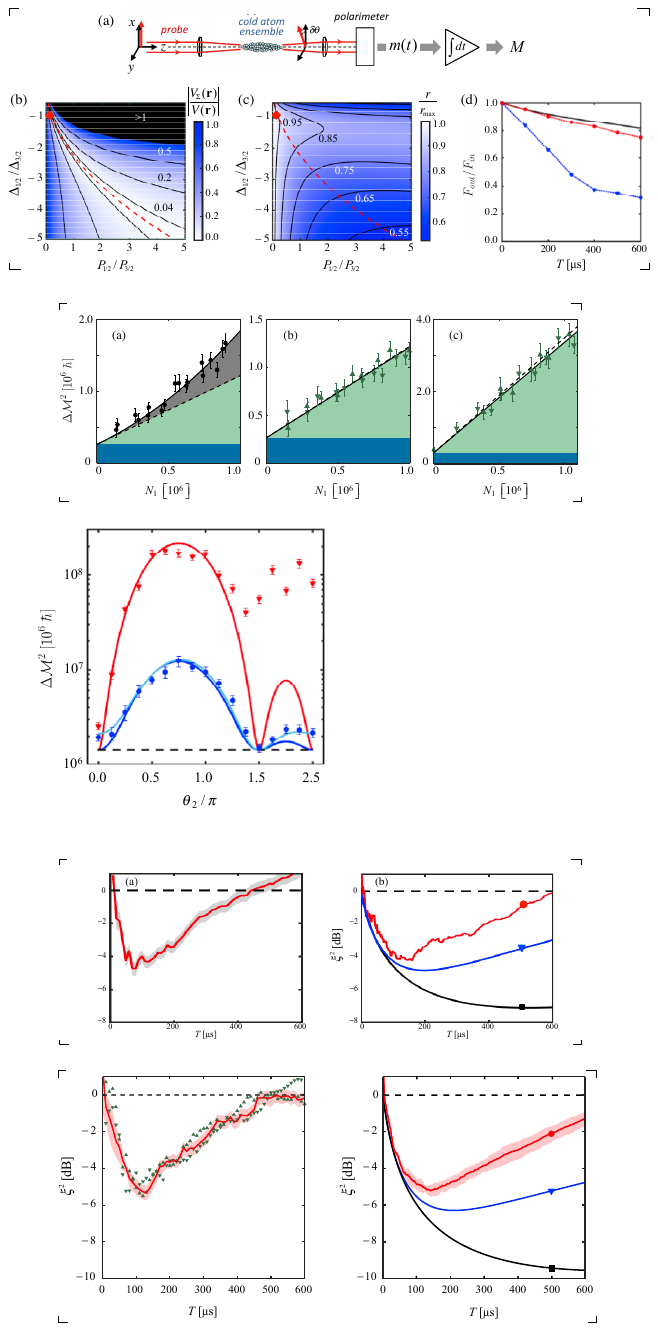}}
\caption{\label{fig:Fig1} (Color online). (a) Experimental setup for optical measurement of a collective atomic spin $\hat{F}_z$. A linearly polarized optical probe does a single pass through a cloud of $\sim 10^6$ Cs atoms held in an optical trap inside a magnetic shield (not shown), and the resulting spin dependent Faraday rotation is measured with a shot-noise limited polarimeter.  The polarimeter output $m(t)$ is integrated for a time $T$ to yield a measurement $M$ of the QDN observable $\hat{M} \propto \hat{F}_z$. (b) Two-color tensor light shift $V_{\Sigma}({\bf r})=V_{1/2}({\bf r})+V_{3/2}({\bf r})$, in units of the tensor light shift $V({\bf r})$ when probing with a single color on the $j'=3/2$ transition and using the same detuning and total power. (c) Measurement strength $r$ for two-color probing, in units of $r_{\text{max}}$ for single-color probing. Dashed red lines indicate where $V_{\Sigma}({\bf r})=0$, and red diamonds indicate the operating point used in the experiment. For the plots in (b) and (c) we set $\Delta_{3/2}=-580\Gamma_{3/2}$, the value used in the experiment. (d) Experimentally observed cancellation of the tensor light shift, showing decay of the mean spin due to optical pumping (black), optical pumping plus tensor light shift (blue triangles), and optical pumping plus two-color cancellation of the tensor light shift (red circles).}
\end{figure*}

\section{Experimental Implementation of QND Measurement}

\noindent A schematic of our experiment is shown in Fig. 1(a).  Our atomic ensemble consists of $\sim 10^6$ Cs atoms prepared in the $6S_{1/2} (f=4)$ hyperfine manifold,  and held in an elongated dipole force trap located inside a high-quality, three-layer magnetic shield that attenuates ambient fields by a factor $\geq 10^4$ in a bandwidth from DC to $50$kHz.  The optical probe is a linearly polarized TEM$_{00}$ spatial mode laser beam doing a single pass though the atom cloud, after which the spin-dependent Faraday rotation is measured with a shot-noise limited polarimeter whose output is integrated for some time $T$. The coupled spin-probe system is governed by a Hamiltonian 
\begin{equation}
\begin{aligned}
\hat{H}=\hbar\Omega\hat{F}_z+\hbar\chi\hat{F}_z\hat{S}_3+\hat{H}^{(2)}
\end{aligned}
\end{equation}
\noindent where $\hat{F}_z$ is in units of $\hbar$, and the three terms represent the interaction with a bias magnetic field along the $z$ axis, the atom-probe Faraday interaction, and a small but important atom-probe irreducible rank-2 tensor interaction (referred to as the ``tensor" interaction in what follows) which we will set aside for now and return to below. The magnetic and Faraday interaction terms commute and the former has no effect on the latter. Accordingly, in what follows we will refer to spin observables in the rotating frame. For a single-color probe tuned near one of the $6S_{1/2}\rightarrow 6P_{j'}$, $j'=\{1/2,3/2\}$  transitions of an alkali atom, this scenario is well studied and we here give only a brief summary of the most important aspects. The reader is referred to the literature  \cite{Deutsch2010, Baragiola2014, Kuzmich1998, Takahashi1999, Kubasik2009} for more detail. 
	
The Faraday interaction $\hbar\chi\hat{F}_z\hat{S}_3$ couples the $3$-component of the probe Stokes vector to a collective spin mode $\hat{F}_z=\sum_n\beta(\mathbf{r}_n)\hat{f}_z^{(n)}$, where the weights $\beta(\mathbf{r}_n)=I(\mathbf{r}_n)/I_{\rm max}$ are given by the intensities seen by the atoms relative to the peak intensity at the probe waist, and where the quantity $N_1=\sum_n\beta(\mathbf{r}_n)$ is an effective atom number defined such that $F=N_1 f$. The strength of the coupling is characterized by the polarization rotation angle $\chi=-C^{(1)}_{j'}(\sigma_0/A)(\Gamma/2\Delta)$, where $\sigma_0=3\lambda^2/2\pi$ is the resonant scattering cross section for unit oscillator strength, $A=\pi w_0^2$ is the characteristic probe cross section, $\Gamma$ is the natural linewidth of the atomic transition, and $\Delta$ is the detuning from resonance. Here and throughout, the notation $C^{(K)}_{j'}$ refers to effective tensor coefficients associated with the atom-probe interaction, see Appendix A for details.

We configure the experiment such that the input Stokes vector lies along the 1-axis of the Poincare sphere and the polarimeter measures $\hat{S}_2$. Note that for a traveling wave probe the Stokes vector components are related to photon flux rather than photon number. Thus, for weak Faraday interaction, the input-output relation is $\hat{S}_2^{out}=\hat{S}_2^{in}+\chi\hat{S}_1^{in}\hat{F}_z$, where $\hat{S}_1^{in}\simeq\dot{N}_{\rm L}/2$  for large $\langle\hat{S}_1^{in}\rangle$, and $\dot{N}_{\rm L}=P/\hbar\omega$ is the photon flux in a probe of power $P$ and frequency $\omega$. The polarimeter output $m(t)$ is a continuous-time measurement of the observable
\begin{equation}
\begin{aligned}
\hat{m}(t)=\hbar\omega(2\hat{S}_2^{in}dt + \chi\dot{N}_{\rm L}T\hat{F}_z), 
\end{aligned}
\end{equation}

\noindent and integrating it for a time $T$ yields a measurement $M$ of our essential QND observable
\begin{equation}
\begin{aligned}
\hat{M}=\hbar\omega(\int_0^T 2\hat{S}_2^{in}dt + \chi\dot{N}_{\rm L}T\hat{F}_z ), 
\end{aligned}
\end{equation}

\noindent where the two parts correspond to contributions from probe shot noise and the Faraday signal, respectively. In principle there is also a contribution to $\hat{m}(t)$ and $\hat{M}$ from the tensor interaction $\hat{H}^{(2)}$, but this is already small for the probe detunings considered here and averages to zero in the rotating frame \cite{Deutsch2010}. Thus, given a measurement outcome $M$, the Maximum Likelihood estimate for the spin is $F_z=M/g(T)$, where $g(T)=\hbar\omega\chi\dot{N}_L T$ is the integrated polarimeter output per unit angular momentum.

The backaction from a measurement of $\hat{M}$ is quantified by the measurement strength 
\begin{equation}
\begin{aligned}
r=\Delta M_{\rm PN}^2/\Delta M_{\rm SN}^2=\chi^2\dot{N}_{\rm L} T\Delta F_z^2 , 
\end{aligned}
\end{equation}

\noindent where $\Delta M_{\rm PN}^2=(\hbar\omega)^2\chi^2\dot{N}_{\rm L}^2 T^2 \Delta F_z^2$ and  $\Delta M_{\rm SN}^2=(\hbar\omega)^2\dot{N}_{\rm L} T$ are the contributions to the overall measurement variance $\Delta M^2$ from spin quantum projection noise (PN) and probe shot noise (SN). In the absence of atom loss and decoherence from optical pumping, measurement backaction would in principle produce a spin squeezed state (SS) with quantum projection noise reduced by a factor $1/(1+r)$.  A more relevant measure is the so-called metrological squeezing, 
\begin{equation}
\begin{aligned}
\xi_{\rm m}^2=(\Delta F_z^2/\langle\hat{F}_x\rangle)_{SS}/(\Delta F_z^2/\langle\hat{F}_x\rangle)_{CS},
\end{aligned}
\end{equation}
\noindent defined as the improvement in sensitivity relative to a spin coherent state (CS) when measuring small rotations \cite{Wineland1994}. Loss and decoherence will affect the quantum projection noise and mean spin on a time scale set by the characteristic photon scattering rate $\gamma$, which  limits the useful measurement time to $T\sim\gamma^{-1}$.  For alkali atoms and large detuning one finds that $\chi^2 \dot{N_{\rm L}}\propto \gamma$, and the minimum value of $\xi_{\rm m}^2$ becomes independent of probe power and detuning, and also whether one probes near the $6S_{1/2} \rightarrow 6P_{j=1/2}$ (D1) or $6S_{1/2} \rightarrow 6P_{j=3/2}$ (D2) transition.

\section{Tensor Light Shifts}

\noindent Beyond loss and decoherence, metrological squeezing in our system is strongly affected by the atom-probe tensor interaction.  Because the tensor polarizability has negligible effect on the probe polarization it can be effectively modeled as a single-atom tensor light shift.  For linear probe polarization along $x$ this takes the form 
\begin{equation}
\begin{aligned}
\hat{H}^{(2)}=\sum_n C^{(2)} V(\mathbf{r}_n) (\hat{f}_x^{(n)})^2, 
\end{aligned}
\end{equation}
\noindent where the scalar magnitude of the light shift, $V(\mathbf{r}_n)$,  depends on the atomic transition, the local probe intensity, and the detuning.  Note that $[\hat{H}^{(2)},\hat{F}_z]\neq 0$, which means the tensor interaction compromises the QND character of a spin measurement via the Faraday interaction.  In \cite{Koschorreck2010a} a QND measurement was recovered, on average, by rapidly alternating the probe between $x$ and $y$ polarization. We similarly recover an (imperfect) QND measurement by adding a bias field along $z$, in which case the effective tensor light shift in the rotating frame takes the form 
\begin{equation}
\begin{aligned}
\hat{H}_{\rm RF}^{(2)}=\sum_n C^{(2)} V(\mathbf{r}_n) (\hat{f}_z^{(n)})^2/2,
\end{aligned}
\end{equation}
\noindent and thus $[\hat{H}_{\rm RF}^{(2)},\hat{F}_z]= 0$  \cite{Smith2004}.  This turns out to be insufficient when working with spin-$f\ge1$  atoms, for which a Hamiltonian of this type will drive complex dynamics of the internal atomic spin \cite{Deutsch2010} in a manner that is inhomogeneous across the ensemble and cannot be undone by standard dynamical decoupling techniques.  The result is a rapid collapse of the mean spin that interferes with the QND measurement and prevents any significant degree of metrological squeezing. 

In our experiment we minimize the overall tensor light shift with a two-color probe consisting of spatially mode matched components near the D1 ($j'=1/2$) and D2 ($j'=3/2$) transitions; from here on we label with $j'$ any quantity that differs between the two transitions and associated probe fields. For alkali atoms at detunings much larger than the excited state hyperfine splitting, we have $\chi_{j'} \propto C_{j'}^{(1)}/\Delta_{j'}$ and $V_{j'}(\mathbf{r}_n)\propto C_{j'}^{(2)}P_{j'}/\Delta_{j'}$, with tensor coefficients \cite{Deutsch2010} \\
\begin{equation}
\begin{aligned}
&C_{1/2}^{(1)} \approx 1/{3f}, \quad &C_{1/2}^{(2)} \approx \zeta_{1/2} \Gamma_{1/2}/\Delta_{1/2}, &\quad  \zeta_{1/2} >0, \nonumber \\ \\
&C_{3/2}^{(1)} \approx -1/{3f},  &C_{3/2}^{(2)} \approx \zeta_{3/2} \Gamma_{3/2}/\Delta_{3/2}, &\quad  \zeta_{3/2} <0. 
\nonumber
\end{aligned}
\end{equation}

\begin{figure*}
[t]\resizebox{18cm}{!}
{\includegraphics{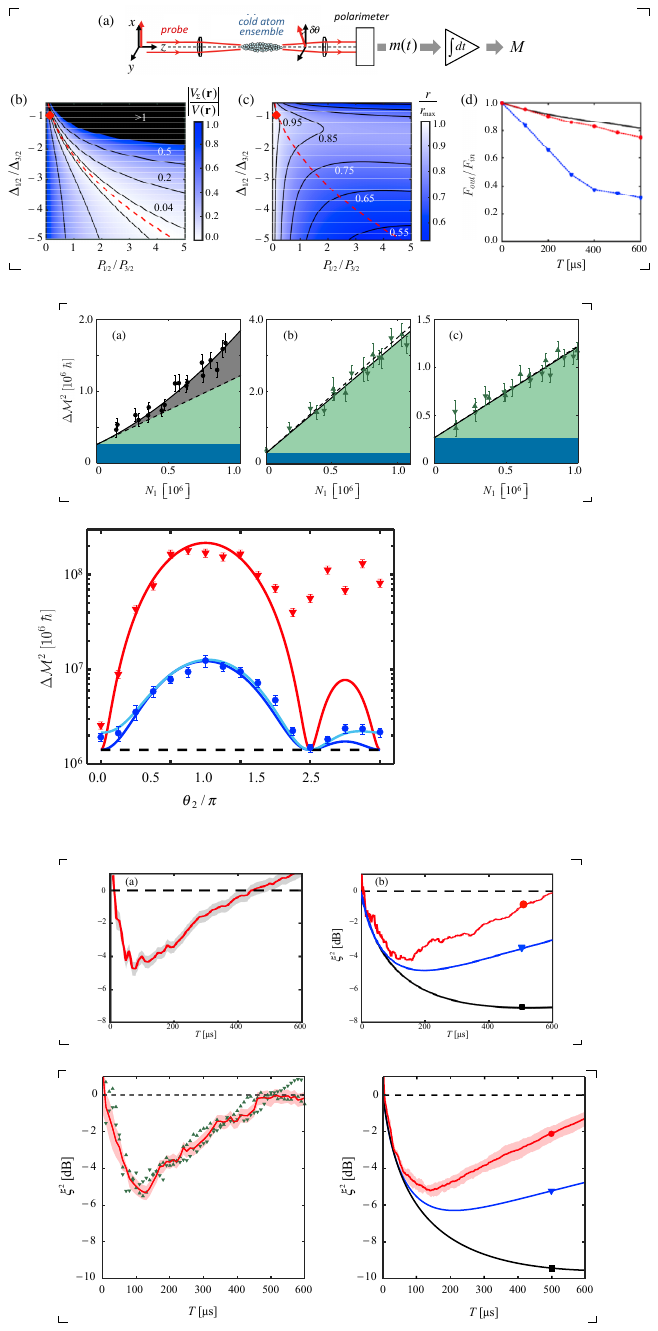}}
\caption{\label{fig:Fig2} (Color online). (a) Measurement variance $\Delta M^2$ for a spin coherent state ($T=120\, \mu$s) versus $N_1$.  The data (black circles) are fitted with a function $\Delta M^2=\Delta M_{SN}^2+aN_1+bN_1^2$ (black line), to separate out probe shot noise (blue), quantum projection noise (green), and classical projection noise (gray).  The dashed line is the predicted $\Delta M_{PN}^2$ based on the known geometry and the measured $N_1$. Preparation of the collective spin includes a noisy rotation by $\pi/2$ around $y$; this is the source of the observed classical projection noise. (b) Two data sets taken on different days (up/down triangles), showing the fitted $\Delta M^2$ (solid line) and predicted $\Delta M_{PN}^2 $(dashed line) for a maximally mixed spin state.  The spherical symmetry of the state eliminates sensitivity to noisy rotations and thus classical projection noise. (c) Two data sets taken on different days (up/down triangles), equivalent to (a) except that a robust composite pulse is used to implement the $\pi/2$ rotation around $y$.  This almost entirely removes classical projection noise, leaving only shot noise and quantum projection noise.  In all cases the measurement variance is estimated from 100 trials at the lowest atom number, gradually increasing to 500 trials at the highest atom number, such that the error bars (standard deviation of the mean) on $\Delta M^2$ remain similar for different $N_1$.}
\end{figure*}

\noindent Thus, when the detunings $\Delta_{1/2}$ and $\Delta_{3/2}$ have opposite sign the Faraday signals add constructively and the tensor light shifts counteract each other. Exact cancellation occurs when $C_{1/2}^{(2)} V_{1/2}(\mathbf{r}_n)=-C_{3/2}^{(2)} V_{3/2}(\mathbf{r}_n)$, which can be achieved for a range of powers and detunings of the $j'=1/2$ and $j'=3/2$ components. The possible combinations can be found using Eq. (A3) in Appendix A, and Fig. 1(b) shows the degree of cancellation as a function of the ratios $\Delta_{1/2}/\Delta_{3/2}$ and $P_{1/2}/P_{3/2}$ over the relevant parameter regime.

Maximizing the measurement strength brings further constraints. The measured observable is now 
\begin{equation}
\begin{aligned}
&\hat{M}=\sum_{j'}\hat{M}_{j'},\qquad \text{where} \\ 
&\hat{M}_{j'}=\hbar\omega_{j'}( \int_0^T 2(\hat{S}_{2}^{in})_{j'} dt + \chi_{j'}\dot{N}_{j'} T \hat{F}_z). 
\end{aligned}
\end{equation}
The variances are $\Delta M_{\rm PN}^2=(\sum_{j'} \hbar \omega_{j'} \chi_{j'}\dot{N}_{j'})^2 T^2 \Delta F_z^2$ and $\Delta M_{\rm SN}^2=(\sum_{j'} \hbar^2 \omega_{j'}^2\dot{N}_{j'}) T$, and the measurement strength is again $r=\Delta M_{\rm PN}^2/\Delta M_{\rm SN}^2$. Finally, in analogy to single-color probing, the metrological squeezing will peak for a measurement time $T \sim (\sum_{j'}\gamma_{j'})^{-1}$. As outlined in Appendix A, it is straightforward to numerically calculate both the measurement strength and the tensor light shift for arbitrary powers and detunings. Experimentally, we have found a sweet spot around $\Delta_{3/2}{=}-580\,\Gamma_{3/2}$ ($-3.0\,$GHz) where both probe absorption and scalar light shifts are negligible. Given this choice, Fig. 1(c) shows a contour plot of the measurement strength $r$ for a two-color probe relative to $r_{\text{max}}$ for a single-color probe, as a function of the ratios $\Delta_{1/2}/\Delta_{3/2}$ and $P_{1/2}/P_{3/2}$. We find a broad maximum near $\Delta_{1/2}/\Delta_{3/2}{=}-1$, where $V_{\Sigma}({\bf r} )\approx 0$ and $r \approx 0.96\, r_{\text{max}}$ for probe powers $P_{1/2}/P_{3/2}\approx 0.2$.  In our experiment, the total probe power is $P_{1/2}+P_{3/2}\approx 20\mu$W, focused to a waist ($1/{\text e}^2$) of $26\mu$m at the center of the atom cloud.

Cancellation of the tensor light shift can be quantified in the experiment by measuring the decay of the mean spin as a function of time. To determine the baseline behavior for single-color probing, we first prepare the collective spin in the state $|F,M_z=F \rangle$ and measure $\hat{F}_z$ as a function of time $T$. Because we start in an eigenstate of the light shift Hamiltonian, any decay of $\langle \hat{F}_z \rangle$ is entirely due to optical pumping which sets the fundamental limit. To determine the worst-case effect of tensor light shifts we next do a variant of the first experiment, in which we rotate the spin to point along $x$, turn on the probe for a time $T$, then rotate the spin back to point along $z$ and immediately measure $\hat{F}_z$. Finally we repeat the second experiment with a two-color probe chosen such that the tensor light shifts cancel. Figure 1(d) shows how this plays out with probe parameters for which peak metrological squeezing occurs with back-to-back measurements of $T \approx 100\, \mu$s each. Notably, at $200\, \mu$s the mean spin has decayed to $92\%$ of its initial value when subject only to optical pumping, to $66\%$ of its initial value when subject to a combination of optical pumping and the tensor light shift in a single-color probe, and to $91\%$ of its initial value when subject to optical pumping and a two-color probe optimized for tensor light shift cancellation.  Operationally, we find a good working point by setting the probe powers and detunings as close as possible to a chosen set of optimal values, and then fine tuning $P_{1/2}$ to minimize the observed spin decay.

\section{Quantum Projection Noise}

\noindent With the two-color Faraday QND measurement in place, the observation of spin quantum projection noise is relatively straightforward.  The basic sequence begins by preparing the individual atomic spins in the state $|f,m_z=f \rangle$, through a combination of optical pumping and selective removal of atoms in other states.  This is equivalent to preparing the collective spin mode in $|F,M_z=F \rangle$, where $F=N_1 f$ and the effective atom number $N_1$ can be found from a measurement of $\hat{F}_z$.  We next apply a single radio-frequency (rf) pulse at the bias Larmor frequency $\Omega=250\,$ kHz to rotate the spin by $\pi/2$ around $y$, resulting in a close-to-minimum-uncertainty spin coherent state that is aligned approximately along $x$.  At this point we turn on the optical probe and record the output of the polarimeter for several ms. Integrating the measurement record from $t=0$ to $T$ then yields a measurement of the observable $\hat{M}$.  Repeating the sequence at least $100$ times allows us to estimate $\Delta M^2$, and doing the same without atoms in the trap gives an independent estimate of $\Delta M_{\rm SN}^2$.  

Figure 2(a) shows a typical data set consisting of estimates for $\Delta M^2$ at $T=120\mu$s, for a range of effective atom numbers $N_1$.  We fit this data with a function
\begin{equation}
\Delta M^2=\Delta M_{\rm SN}^2+aN_1+b(N_1)^2,
\end{equation}

\begin{figure}
[t]\resizebox{8.5cm}{!}
{\includegraphics{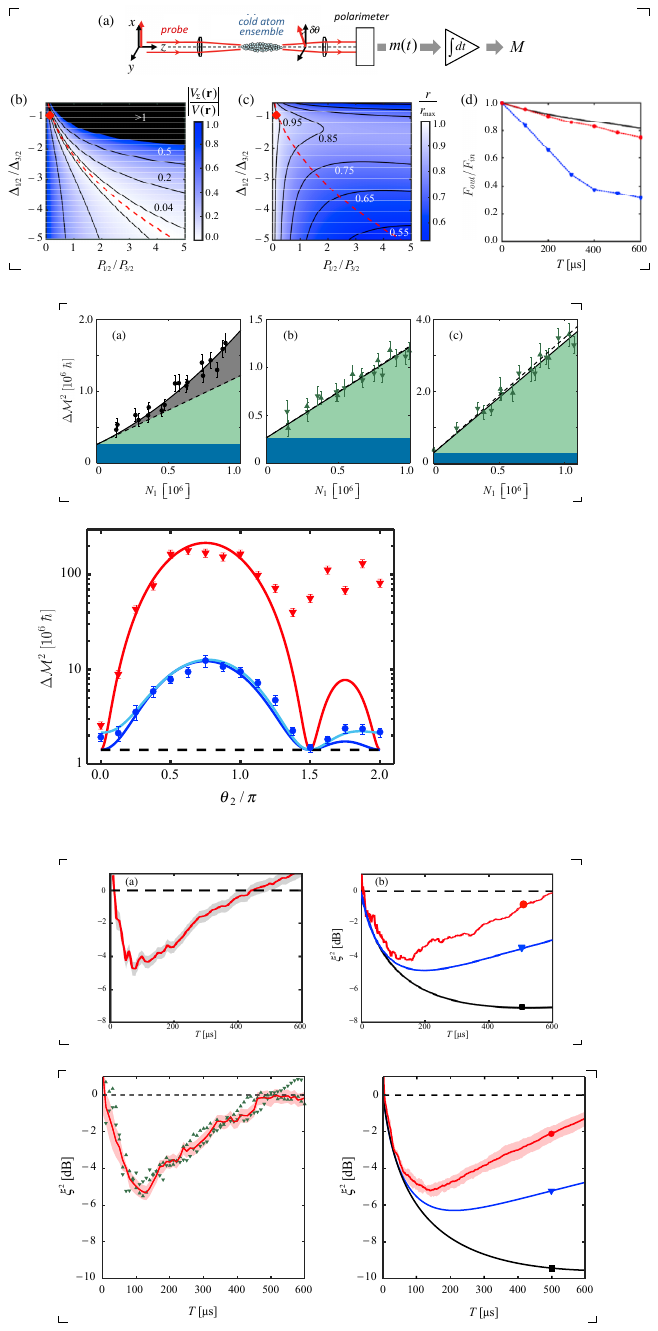}}
\caption{\label{fig:Fig3} (Color online) Initial state preparation with a composite rotation $R(\theta_2) = R(\theta_2,\epsilon_2)R(\theta_1,\epsilon_1)$, where nominally $\theta_1 = \pi/2$, $\epsilon_1 = y$, $ \epsilon_2=x$, and $\Delta M^2$ is measured as function of  $\theta_2$.  For $\theta_2=3\pi/2$ the sequence is robust to fluctuations in both amplitude and detuning of the rf field that drives the rotations; for $\theta_2=0, 2\pi$ it is robust to detuning but not amplitude fluctuations.  Blue dots are experimental data, the dark blue line is the predictions of a model with detuning fluctuations only, and the light blue line is the prediction of a model with detuning and amplitude fluctuations. Red triangles are data taken in the absence of magnetic shielding, showing a poor fit to the model due to ambient AC magnetic fields}
\end{figure}

\noindent where the three terms correspond to probe shot-noise, quantum projection noise, and ``classical" projection noise resulting from errors in the rotation that puts the spin-coherent state along $x$.  Focusing on the quantum projection noise, $\Delta M_{\rm PN}^2$, we find that for this data set it exceeds $\Delta M_{\rm SN}^2$  by $5.1$ dB at the largest $N_1$, corresponding to a measurement strength $r=3.25$.  In the absence of loss and decoherence, this value implies a post-measurement reduction of projection noise (quantum and classical combined) to a level $6.3$ dB below the spin coherent state.  However, this is not a good approximation given the characteristic photon scattering rate, $1/\gamma \sim 35\,\mu$s, for this data set.  Also shown (Fig. 2b) is a data set where the collective spin is initially prepared in a maximally mixed (thermal) state, for which the quantum projection noise is a factor of $10/3$ larger than for the spin coherent state. The spherical symmetry of this state means it is not affected by noisy rotations or tensor light shifts during the measurement, and thus rules out any introduction of classical projection noise. The maximally mixed state is prepared as in \cite{Koschorreck2010b}, and serves as a very robust calibration of the quantum projection noise present in our experiment. Separately, the quantum projection noise observed in Figs. 2(a) and (b) are in very close agreement with a prediction based on the measured $N_1$ and the known geometry of the experiment.

\begin{figure*}
[t]\resizebox{18cm}{!}
{\includegraphics{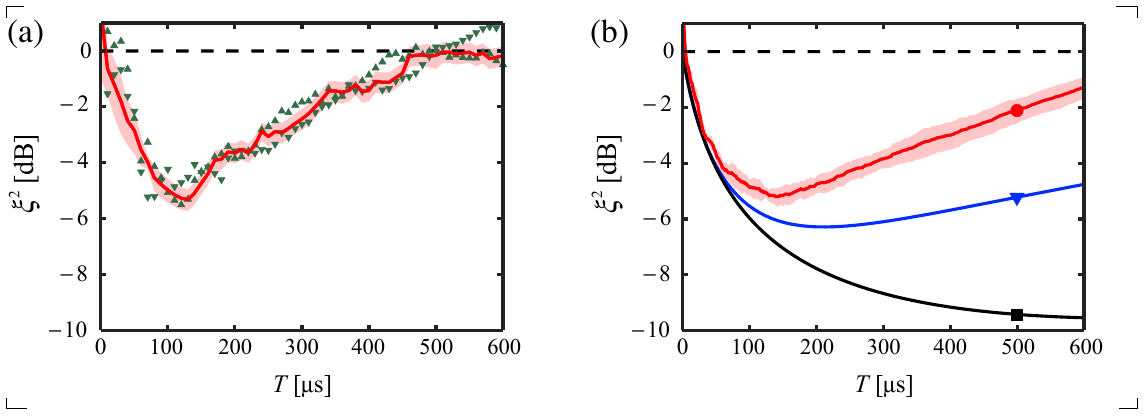}}
\caption{\label{fig:Fig4} (Color online) (a) Metrological squeezing parameter $\xi^2$, for $N_1=10^6$ and estimated from 2,700 back-to-back measurements of duration $T$. Up/down triangles correspond to two separate data sets taken on different days; the solid red line is the estimated metrological squeezing when combining both sets and analyzing them as one, and the shaded area is the corresponding one standard deviation error band. (b) Results from a detailed theoretical model of spin squeezing in our experiment, showing the predicted reduction in quantum projection noise (black line/square) and the predicted metrological squeezing $\xi_m^2$ defined as in Eq. 5 (blue line/triangle); these quantities are not accessible in the experiment.  The red line/dot shows the predicted metrological squeezing $\xi^2$, defined as in Eq. 10, and estimated from 27,000 simulated measurement records that have been analyzed using the exact same protocol as measurement records from the experiment.  A peak metrological squeezing of $\gtrsim -5$dB is observed in both the experiment and simulated data.}
\end{figure*}

\section{Composite Pulse Control}

\noindent The classical projection noise visible in Fig. 2(a) is a stark reminder of the challenge posed by dynamical control of large collective spins. As already outlined, for this data set a single rf pulse was applied to drive a rotation $R(\theta_1,\epsilon_1)$, where nominally $\theta_1{=}\pi/2$ and $\epsilon_1{=}y$.  In practice, both the angle and axis of rotation is fluctuating due to variations in rf amplitude and detuning.  Thus, if $\theta_1$ is subject to Gaussian fluctuations with zero mean and variance $\Delta\theta_1^2$, classical and quantum projection noise will be equal when $\Delta\theta_1= \Delta F_z/F$.  That is, for a collective spin $F\sim 10^6$, the rms angle error must be of order $10 \mu$rad or less for classical projection noise to be negligible. Furthermore, fluctuations in the bias magnetic field affects the detuning and thus the axis of rotation, $\epsilon_1=y+(\Delta\Omega/\Omega_{\text {rf}})z$, where $\Delta \Omega$ is the rms error in the bias Larmor frequency and $\Omega_{\text {rf}}$ is the rf Larmor frequency. In the absence of magnetic shielding, this is the dominant source of classical projection noise in our experiment. 

To overcome this problem, we use a composite rotation implemented with two consecutive rf pulses, $R(\theta_2) = R(\theta_2,\epsilon_2)R(\theta_1,\epsilon_1)$, where nominally  $\epsilon_1=y$, $\theta_1=\pi/2$, $\epsilon_2=x$, and the two pulses are subject to identical amplitude and detuning errors.  As seen in Fig. 3, the overall projection noise $(\Delta F_z)^2$ is insensitive to both amplitude and detuning errors at $\theta_2=3\pi/2$, and to detuning errors only at $\theta_2=0, 2\pi$. Data taken for a range of $\theta_2$ are in good agreement with predictions assuming Gaussian amplitude and detuning variations from run to run of our experiment, and reaches the quantum projection noise floor at $\theta_2=3\pi/2$.  This is the operating point used for the data in Fig. 2(c), showing that we can reliably reach the quantum projection noise floor for effective atom numbers in the entire range $N_1 \leq 10^6$.  A separate set of data in Fig. 3 shows that our composite pulse performs poorly in the absence of magnetic shielding, most likely because the detuning error is changing during the $160\mu$s pulse duration.  This is consistent with the presence of residual magnetic fields in the DC-$50$kHz range in our laboratory.

\section{Metrological Spin Squeezing}

\noindent Squeezing by measurement backaction is inherently non-deterministic, and applying it to metrology or sensing requires two successive measurements, one before and one after the spin-rotation of interest. Thus, the appropriate measure of metrological advantage is based on the conditional variance of back-to-back measurements $\hat{M}_1$  and $\hat{M}_2$ \cite{Sewell2012},
\begin{equation}
\begin{aligned}
\xi^2=\frac{\Delta(M_2|M_1)^2-\Delta M_{\rm SN}^2}{\Delta M_{\rm PN}^2} \frac{|\langle \mathbf{\hat{\mathbf F}}_2 \rangle |^2}{|\langle {\hat{\mathbf F}}_1 \rangle |^2} 
\label{eq:eq1}
\end{aligned}
\end{equation}

\noindent As defined, this metrological squeezing parameter accounts for squeezing of the conditional variance, as well as the injection of quantum projection noise and loss of mean spin that occurs due to atom loss and optical pumping. To estimate the  conditional variance from experimental data, we first relate it to the covariance, $\Delta (M_2|M_1)^2=\Delta M_2^2-\text{cov}(M_1,M_2)^2/\Delta M_1^2$, and then estimate $\text{cov}(M_1,M_2)=[\Delta(M_1+M_2)^2-\Delta(M_1-M_2)^2]/4$ directly from sets of measurement records that we separate into two sections of length $T$.  Finally, the reduction in mean spin during the first measurement is obtained from the data in Fig. 2.  Figure 4 is a typical data set showing $\xi^2$ versus $T$, attaining a metrological squeezing of $5$ dB to $5.5$ dB at just above $100\,\mu $s.

\section{Comparison to theory}
\noindent Our experiment offers a unique opportunity to validate the predictions of a full, quantitative model of Faraday interaction-based quantum measurements of collective atomic spins, as shown in Fig. 4.  We employ a first-principles stochastic master equation (SME) that includes both the effect of measurement backaction and  a complete description of optical pumping~\cite{Baragiola2014, NorrisThesis2014}.  Such a model is faced with three complicating factors: (i) complex many-body dynamics; (ii) spatial inhomogeneities in the probe intensity and atomic density; and (iii) the complex internal hyperfine levels.  Given the modest coupling strength achieved here,  the many-body state is well approximated as a Gaussian state, described simply by a set of one- and two-body correlations \cite{Kurucz:2010aa, NorrisThesis2014, Qi2018}, the former describing the mean-spin and the latter the spin-spin correlations.  We address inhomogeneity as in~\cite{Baragiola2014} by projecting into a basis of Laguerre-Gaussian and longitudinal spatial modes.   Finally as in~\cite{NorrisThesis2014, Qi2018} we restrict the internal dynamics by treating each atom as a qutrit, ${\ket{\uparrow}, \ket{\downarrow}, \ket{T} }$.   For the initial spin coherent state preparation we have $\ket{\uparrow } = \ket{ f=4, m_{x}=4 }$, and the Faraday interaction leads to entanglement dominantly with $\ket{\downarrow} = \ket{f=4,m_x=3}$.  Optical pumping also couples $\ket{\uparrow } \rightarrow \ket{\downarrow }$ and $\ket{\downarrow }\rightarrow \ket{T} = \ket{f=4,m_x=2}$; the latter is important to account for transfer of coherence~\cite{Norris2012}.  As described in~\cite{NorrisThesis2014, Qi2018} and Appendix B, we use the SME to obtain a closed set of one- and two-body correlations functions in this truncated basis, projected into the basis of spatial modes. These are numerically integrated to generate both a direct theoretical prediction of the spin squeezing parameters as a function of time, and a simulation of the noisy polarimeter signal $m(t)$. When analyzed the same way as experimental data, the latter predicts a degree of metrological squeezing in good quantitative agreement with the experiment, given realistic uncertainties about the atom cloud size and probe geometry.

\section{IV. Summary and Outlook}  

\noindent With the work reported here, we have demonstrated the viability of spin squeezing for large collective spins composed of the individual physical angular momenta of up to $10^6$ Cs atoms. A peak metrological squeezing  of $\geq 5$dB was generated by quantum backaction from a dispersive QND measurement, implemented with a two-color optical probe that minimizes tensor light shifts without sacrificing measurement strength.  Other significant developments include the successful application of composite pulse techniques for accurate dynamical control of collective spins, enabled in part by broadband suppression of background magnetic fields inside a high quality magnetic shield.  Finally, we have compared the observed quantum projection noise and conditional squeezing to a theoretical model that accounts for the relevant atomic physics and collective spin mode, finding good quantitative agreement and validating its use when modeling other, more complex experiments.

Looking ahead, a clear next step is to demonstrate that the observed conditional squeezing conveys an actual metrological advantage when detecting a small spin rotation inserted between back-to-back measurements, both in broadband AC magnetometry near the bias Larmor frequency, and in near-DC magnetometry without a bias field. Having solved the twin problems of tensor light shifts and accurate dynamical control, there are additional avenues to explore, notably the combination of internal and collective spin control. Prior theoretical work from our collaboration suggests that significant additional squeezing and as much as 10 dB improvement in the sensitivity of atomic magnetometry might be achieved, given sufficiently accurate control of the internal atomic state ~\cite{Norris2012}. Other opportunities include a revisit of long-standing ideas related to continuous measurement and real-time feedback for deterministic squeezing and closed-loop magnetometry  \cite{Thomsen2002, Stockton2004}. The introduction of real-time feedback also opens the door to new types of quantum nonlinear dynamics, wherein the evolution of a quantum system is conditioned on the outcome of a time-continuous QND measurement. This has the potential to access new regimes of quantum simulation near the quantum-classical boundary, and opportunities to study long-standing issues related to quantum-classical correspondence in chaotic systems \cite{MunozArias2020a, MunozArias2020b}.

\begin{acknowledgments}
This work was supported by the US National Science Foundation Grants PHY-1306171, PHY-1606989, and PHY-1607125. 
\end{acknowledgments}

\appendix

\section{Quantities that depend on the atomic tensor polarizability}

\noindent The main part of this article uses a compact notation to account for hyperfine structure in the Cs $6P_{1/2}$ and $6P_{3/2}$ excited states.  In the following we summarize the relevant atomic physics ~\cite{Deutsch2010} and define a number of parameters referred to above, e.g., Faraday rotation angles, light shifts, scattering rates, etc. \\

\subsection{Atom-light interaction}

The atom-laser light-shift interaction is characterized by a series of effective tensor coefficients $C_{j'}^{(K)}$, for irreducible rank $K=0,1,2$ scalar, vector, and tensor interactions, respectively. We define

\begin{equation}
C_{j'}^{(K)}=\sum_{f'} C_{j'f'f}^{(K)} \frac{\Delta_{j'}}{\Delta_{j'f'f}},
\end{equation}

\noindent where primed and unprimed quantum numbers refer to electronic excited and ground states, respectively. The individual detunings  $\Delta_{j'f'f} \coloneqq \omega_{j'} - (E_{j'f'} - E_{jf})/\hbar$  are with respect to the $\ket{6S_{1/2}(f)}\rightarrow \ket{6P_{j'}(f')}$ transitions, and we define the effective detuning $\Delta_{j'}$ with respect to the $\ket{6S_{1/2}(f=4)}\rightarrow \ket{6P_{j'}(f_{\rm max})}$ transition.  Expressions for the coefficients $C_{j'f'f}^{(K)}$ can be found in Appendix A of reference \cite{Deutsch2010}.

The vector light-shift leads to Faraday interaction, where we have defined the characteristic Faraday rotation angle as
\begin{align}
\label{chi}
\chi_{j'} =\sum_{f'}\chi_{j'f'}&=\sum_{f'}-C_{j'f'f}^{(1)}\frac{\sigma_{j'}}{A}\frac{\Gamma_{j'}}{2\Delta_{j'f'f}}\\
&=-C_{j'}^{(1)} \frac{\sigma_{j'}}{A} \frac{\Gamma_{j'}}{2\Delta_{j'}} \nonumber.
\end{align}
Here $\Gamma_{j'} =\frac{4}{3\hbar} d_{j'j}^2\frac{ \omega_{j'}^3}{c^3}$ is the characteristic spontaneous emission rate on the $j' \rightarrow j$ transition, $d_{j'j}$ is the reduced dipole moment, and $\sigma_{j'} = \frac{3 \lambda_{j'}^2}{2\pi}$ is the resonant scattering cross section.  

For $x$-polarized light, the irreducible rank-2 tensor light shift for an atom at position $\mathbf{r}_n$, where the local laser electric-field amplitude is $\mathcal{E}_{{\rm L},j'}(\mathbf{r})$ and intensity $I_{j'}(\mathbf{r}_n)$, is
\begin{align}
\label{tensorLS}
\hat{H}_{j'}^{(2)}(\mathbf{r}_n) &= \sum_{f'} C_{j'f'f}^{(2)}\; \frac{\hbar \Omega_{j'}^2(\mathbf{r}_n)}{4 \Delta_{j'f'f}} \left[ \hat{f}^{(n)}_{x} \right]^2 \\
&=\sum_{f'} C_{j'f'f}^{(2)}\frac{\Delta_{j'}}{\Delta_{j'f'f}}\frac{\hbar\Gamma_{j'}}{8}\frac{I(\mathbf{r}_n)}{I_{{\rm sat}, j'}}\frac{\Gamma_{j'}}{\Delta_{j'}} \left[ \hat{f}^{(n)}_{x} \right]^2 \nonumber  \\
&= C_{j'}^{(2)} V_{j'}(\mathbf{r}_n) \left[ \hat{f}^{(n)}_{x} \right]^2,  \nonumber  
\end{align}

\noindent where $\Omega_{j'}(\mathbf{r}_n) = - d_{j'j} \mathcal{E}_{{\rm L},j'}(\mathbf{r}_n)/\hbar$ is the Rabi frequency, and $I_{{\rm sat}, j'}$ is the saturation intensity for unit oscillator strength.

The characteristic photon scattering rate is
\begin{align}
\gamma_{j'} &= \sum_{f'} \frac{\sigma_{j'} I_{{\rm max},j'} }{\hbar\omega_{j'}}\frac{\Gamma_{j'}^2}{4 \Delta_{j'f'f}^2}  \\
&=\frac{\sigma_{j'} I_{{\rm max},j'} }{\hbar\omega_{j'}}\frac{\Gamma_{j'}^2}{4 \Delta_{j'}^2}\sum_{f'} \frac{\Delta_{j'}^2}{\Delta_{j'f'f}^2} \nonumber , 
\end{align}

\noindent where $\sum_{f'} \frac{\Delta_{j'}^2}{\Delta_{j'f'f}^2}\approx1$ for detunings much larger than the excited state hyperfine splitting.

\subsection{Measurement }

For a single-color probe tuned near the $\ket{6S_{1/2}(f=4)}\rightarrow \ket{6P_{j'}}$ transition the measurement strength is 

\begin{align}
r_{j'} &= \frac{(\Delta M_{PN}^2)_{j'}}{(\Delta M_{SN}^2)_{j'}}=\frac{(\hbar\omega_{j'}\chi_{j'}\dot{N}_{j'})^2T_{j'}^2\Delta F_z^2}{(\hbar\omega_{j'})^2\dot{N}_{j'}T_{j'}}  \\
&=\frac{\eta_{j'}^2T_{j'}}{\kappa_{j'}}\Delta F_z^2 \nonumber , 
\end{align}

\noindent where $\dot{N}_{j'}=P_{j'}/\hbar \omega_{j'}$ is the photon flux, and we have defined $\eta_{j'} \coloneqq \hbar\omega_{j'}\chi_{j'}\dot{N}_{j'}$ and $\kappa_{j'}\coloneqq (\hbar\omega_{j'})^2\dot{N}_{j'}$.  Choosing $T_{j'}\sim1/\gamma_{j'}$ we get a measure for the largest useful measurement strength,

\begin{equation}
r_{\rm max}=\frac{\eta_{j'}^2\Delta F_z^2}{\kappa_{j'}\gamma_{j'}}.
\end{equation}

Considering now a two-color probe with components close to the $\ket{6S_{1/2}(f=4)}\rightarrow\ket{6P_{1/2}}$ and $\ket{6S_{1/2}(f=4)}\rightarrow\ket{6P_{3/2}}$ transitions, respectively, we have 

\begin{align}
r &= \frac{(\eta_{1/2}+\eta_{3/2})^2T_{2 {\rm C}}\Delta F_z^2}{\kappa_{1/2}+\kappa_{3/2}}  \\
&= \frac{\eta_{3/2}^2T_{2 {\rm C}}\Delta F_z^2}{\kappa_{3/2}} \frac{(1+\eta_{1/2}/ \eta_{3/2})^2}{1+\kappa_{1/2}/\kappa_{3/2}}. \nonumber 
\end{align}

\noindent In this case the useful measurement time is given by the combined scattering rates, $T_{2 {\rm C}}\sim1/(\gamma_{1/2}+\gamma_{3/2})$.  Substituting and rearranging, we get 

\begin{align}
\frac{\eta_{3/2}^2T_{2 {\rm C}}\Delta F_z^2}{\kappa_{3/2}}&=\frac{\eta_{3/2}^2T_{3/2}\Delta F_z^2}{\kappa_{3/2}}\,\,\frac{T_{2 {\rm C}}}{T_{3/2}} \\
&=r_{\rm max}\,\,\frac{\gamma_{3/2}}{\gamma_{1/2}+\gamma_{3/2}}, 
\nonumber
\end{align}

\noindent and finally an expression for the useful measurement strength in units of its maximum value, 

\begin{equation}
\frac{r}{r_{\rm max}}=\frac{1}{1+\gamma_{1/2}/ \gamma_{3/2}} \frac{(1+\eta_{1/2}/ \eta_{3/2})^2}{1+\kappa_{1/2}/\kappa_{3/2}}.
\end{equation}

Using the effective Faraday rotation angles and scattering rates defined above, it is then straightforward to numerically calculate $r/r_{\rm max}$.  Fig. 1(c) in the main text shows how this quantity varies as a function of the relative probe detunings and powers, for $\Delta_{3/2}=-580\Gamma_{3/2}$.

\section{Three-dimensional model for two-color polarimetry}

We model the spin squeezing in our experiment via a first-principles stochastic master equation that accounts for both measurement backaction and optical pumping in the interaction of the laser beam with the atomic cloud, followed by detection in the polarimeter.  This model builds on the work of Norris and Baragiola~\cite{NorrisThesis2014, Baragiola2014}, and we refer the reader to previous references for much of the detail.   Such a model is faced with three complicating factors: (i) complex many-body dynamics; (ii) spatial inhomogeneities in the probe intensity and atomic density; and (iii) the complex internal structure of hyperfine levels.  We tackle (i) using the Gaussian approximation, tracking only one and two-point correlation functions, which is an excellent approximation in the weak coupling regime.  In~\cite{Baragiola2014} (ii) was addressed by incorporating the three-dimensional character of the cloud and the light, which results in the inhomogeneous scattering of the light into a superposition of transverse modes.  In~\cite{NorrisThesis2014} (iii) was addressed by the using of encoding into qutrits, which captures the essential quantum correlations.  The model in (iii) employed a simplified model of decoherence, applicable when the probe detuning is large compared to the excited state hyperfine splitting.  Here we unify (i), (ii), and (iii), and include a full decoherence model, appropriate for our two-color probe geometry.  The model described here includes all the details on the relevant atomic physics, local and collective decoherences, continuous measurement effects, and the inhomogeneities in the atomic cloud and the probe, and it gives a reliable description of the experimental spin squeezing generated at early times.

\subsection{Light-shift interaction}
For low saturation, the atom-light interaction is described by the light shift Hamiltonian which can be decomposed into irreducible scalar, vector, and tensor parts~\cite{Deutsch2010}. The scalar interaction does not drive dynamics, and the irreducible tensor interaction, $H^{(2)}$ in Eq. (\ref{tensorLS}), is cancelled using two probe lasers, one detuned near the D1-line ($6S{1/2}\rightarrow 6P_{1/2}$ transition) and the other near the D2-line ($6S{1/2}\rightarrow 6P_{3/2}$ transition), as described in the main text. The remaining nontrivial term is the  irreducible vector light-shift. Taking the direction of propagation of the laser by as the $z$-direction,  the vector light-shift couples each the atomic spin operator $\hat{f}_z$ to the $\hat{S}_3$ Stokes component of the field giving to the Faraday interaction.  For the two-color laser probes	\begin{equation}
		\hat{H}_{\rm Faraday} = \chi_{1/2}\hat{f}_z \hat{S}_3 + \chi_{3/2}  \hat{f}_z \hat{S}_3,
	\end{equation}
where $\chi_{j'}$ is given in Eq. (\ref{chi}). Here and throughout we assume that the atom is prepared in the $f=4$ hyperfine manifold of the $6S_{1/2}$ electronic ground state. 

In the presence of the probe beams each atom also undergoes optical pumping.  We track optical pumping with magnetic sublevels in the  $f=4$ manifold, which dominantly couple to the probe; optical pumping into the $f=3$ hyperfine manifold is treated as loss.  The dynamical map that describes optical pumping for a single atom is then~\cite{Deutsch2010}
\begin{align} \label{eq:opmap}
		 \gamma_{j'}(\mathbf{r}) \mathcal{D}_{j'}[\hat{\rho}] \coloneqq  - \frac{i}{\hbar} [\hat{H}^{j'}_{\rm loss} \hat{\rho} - \hat{\rho} \hat{H}^{j' \dag}_{\rm loss} ] + \Gamma_{j'} \sum_q \hat{W}^{j'}_q \hat{\rho} \hat{W}^{j'\dagger}_q,
\end{align}
where $\hat{H}^{j'}_{\rm loss}$ is the anti-Hermitian part of the light shift leading to absorption,  and the jump operators are given as
\begin{equation}
	\hat{W}^{j'}_{q} = \sum\limits_{f'} \frac{\Omega_{j'}(\mathbf{r}) /2}{\Delta_{j',f'f=4} + i \Gamma_{j'} /2} \left( \bold{e}_{q}^{*} \cdot \hat{\bold{D}}^{j'}_{f=4,f'} \right) \left( \vec{\epsilon}_{L} \cdot \hat{\bold{D}}^{j' \dagger}_{f'f=4}\right),
\end{equation}
which arise from absorption of a laser photon with polarization $\vec{\epsilon}_{L}$ and emission of a photon with polarization $q$ (labeling elements of the spherical basis).  The operators $\hat{\bold{D}}^{j' \dagger}_{f'f}$ are dimensionless raising operators from ground to excited sublevels, defined in~\cite{Deutsch2010}.  The local photon scattering rate for a laser tuned near the $j'$ resonance for a unit Clebsch-Gordon coefficient strength is $\gamma_{j'}(\mathbf{r}) = [I(\mathbf{r})/I_{j',{\rm max}}]\gamma_{j'} $.

In the experiment a strong bias magnetic field is applied along the probe propagation direction.  Since the Larmor precession frequency is large compared to the scattering rate, we make a rotating-wave approximation by performing the average over a Larmor cycle,
$\mathcal{D}_{j'}[\hat{\rho}] \rightarrow \tfrac{1}{2 \pi} \int_{\phi = 0}^{2 \pi} d\phi \, \hat{U}(\phi) \,\mathcal{D}_{j'}[\hat{\rho}] \hat{U}^\dagger(\phi)$ where $\hat{U}(\phi) = \text{exp} (-i \phi  \hat{f}_{z})$. 

\subsection{Geometry of the atomic cloud and probe light}

Atoms in the cloud experience a Faraday interaction and optical pumping proportional to the local intensity, which varies across the ensemble according to both the cloud density and the spatially varying probe.
We account for spatial inhomogeneities by projecting into a basis of spatial modes, using an extended version of the model introduced in \cite{Baragiola2014}. 

The distribution of atoms in the optical trap is described by a Gaussian density 
\begin{equation}
\begin{aligned}
\eta({\bold{r}}) = \eta_{0} \hspace{1mm} \text{exp} \left(  -2 \frac{ \mathbf{r}_\perp ^{2}}{w ^{2}_{\perp}} -2 \frac{z^{2}}{w ^{2}_{z}} \right)
\end{aligned}
\end{equation}
where $w_{\perp}^2$ and $w_{z}^2$ are $e^{-2}$ variances along the perpendicular and parallel directions to the probe, respectively, and $\eta _{0}$ is the peak density at the center of the cloud. The total number of atoms is $N_A = \int dz \, d^2 \mathbf{r}_\perp \, \eta({\bold{r}})$.

Each probe is a linearly polarized, paraxial TEM$_{00}$ beam with electric field $\vec{\mathcal{E}}_{L,j'}(\mathbf{r}_\perp,z) = \mathbf{e}_x \mathcal{E}_{L,j'} u_{00}(\mathbf{r}_\perp,z)$, where $\mathbf{e}_x$ is the probe polarization, $\mathcal{E}_{L,j'}$ is the peak electric-field amplitude, and $u_{00}(\mathbf{r}_\perp,z)$ is the fundamental Laguerre-Gauss (LG) spatial mode. The transverse LG spatial modes are dimensionless and orthonormal: $\int d^2 \mathbf{r}_\perp u^*_{pl}(\mathbf{r}_\perp,z) u_{p'l'}(\mathbf{r}_\perp,z) = A \delta_{p,p'} \delta_{l,l'}$, where $A$ is the transverse beam area. 

We decompose the probe light collectively scattered by the atoms into the basis of transverse LG modes. 
Each paraxial mode of light in the far field (at the polarimeter) is coupled via the Faraday interaction to a collective spin wave defined across the atomic ensemble,
\begin{align}
	\hat{F}_z^{pl} = \sum_{n=1}^{N_A} \beta_{pl}(\mathbf{r}_n) \hat{f}_z^{(n)}.
\end{align}
Here, $\mathbf{r}_n$ is the position of the atom, and the complex-valued weighting coefficients, $\beta_{pl}(\mathbf{r}) = u^*_{pl}(\mathbf{r}_\perp,z) u_{00}(\mathbf{r}_\perp,z),$ can be interpreted as the absorption of an $\mathbf{e}_x$-polarized probe photon in the fundamental $00$-mode followed by emission of a $\mathbf{e}_y$-polarized photon into the $pl$-mode. The polarimeter measures in the $45^\circ$ polarization basis, selecting the $\mathbf{e}_y$-polarized scattered light in the fundamental $00$-mode (the Faraday rotation signal) via homodyne-type detection, with the large-amplitude probe serving as the local oscillator. The result is that the polarimeter performs an effective measurement of the fundamental spin wave $\hat{F}_z^{00}$, referred to in the main text for brevity as $\hat{F}_z$.

\subsection{Stochastic master equation}

In the infinitesimal limit the differential signal from the polarimeter is described by
	\begin{equation} \label{eq:polarimetersignal}
		dM = \langle \hat{F}_{z}^{00} \rangle dt + \tfrac{1}{\sqrt{\kappa}} dW,
	\end{equation}
where the first term is the Faraday rotation signal from the light scattered into the spatial mode of the probe, and the second is the shot noise from the probe described by the Wiener process $dW$. 

The continuously monitored polarimetry signal can be used to provide a conditional update of the collective atomic state using a stochastic master equation.
For two-color probing, the dynamics of the collective atomic state, $\hat{\rho}_C$, is governed by the stochastic master equation \cite{Baragiola2014},
	\begin{align}\label{SME}
d \hat{\rho}_C = & \sqrt{\frac{\kappa}{4}} \mathcal{H}_{00} [\rhocoll] dW + \frac{\kappa}{4} \sum\limits_{p,l} \mathcal{L}_{pl}[\rhocoll] dt \\
& +\sum_{j'} \sum_{n=0}^{N_A} \gamma_{j'}(\bold{r}_{n}) \mathcal{D}^{(n)}_{j'} [\rhocoll] dt. \nonumber 
	\end{align}
The conditional dynamics from the continuous polarimetry measurements are described by the first two terms, with the superoperators defined as
\begin{align}
& \mathcal{H}_{pl}[\rhocoll]  \coloneqq \hat{F}_{z}^{pl} \rhocoll +  \rhocoll \hat{F}_{z}^{pl \dagger} - \text{Tr} [ (\hat{F}_{z}^{pl} + \hat{F}_{z}^{pl \dagger}) \rhocoll ] \rhocoll , \\
& \mathcal{L}_{pl}[\rhocoll] \coloneqq \hat{F}_{z}^{pl} \rhocoll \hat{F}_{z}^{pl \dagger} - \tfrac{1}{2} \hat{F}_{z}^{pl \dagger} \hat{F}_{z}^{pl} \rhocoll - \tfrac{1}{2} \rhocoll  \hat{F}_{z}^{pl \dagger}\hat{F}_{z}^{pl} . 
\end{align}
The first term, $\mathcal{H}_{pl}[\rhocoll] $, drives the conditional dynamics that depend on the measurement signal, and the $ \mathcal{L}_{00}[\rhocoll]$ term describes the associated backaction. The other terms, $ \mathcal{L}_{pl}[\rhocoll]$ for $pl \neq 00$, give the additional collective decoherence from unmeasured forward-scattered light in other spatial modes.  The effective measurement rate
	\begin{align}
		\kappa = \frac{  \left[ \text{sgn}(\chi_{1/2}) \sqrt{ \dot{N}_{1/2} \kappa_{1/2}} + \text{sgn}(\chi_{3/2}) \sqrt{ \dot{N}_{3/2} \kappa_{3/2}} \right]^2 }{   \dot{N}_{1/2} +  \dot{N}_{3/2}  }
	\end{align}
is composed from the measurement rates for each probe,
	\begin{align}
		\kappa_{j'} = \chi_{j'}^2 \dot{N}_{j'},
	\end{align}
where $\dot{N}_{j'}$ is the photon flux in the probe tuned near the $\ket{6P_{j'}}$ resonance.

The final term in the SME describes the effects of \emph{local} optical pumping as individual atoms diffusely scatter photons proportional to their local scattering rate $\gamma_{j'}(\mathbf{r}) = [I_{j'}(\mathbf{r}) /I_{j',{\rm max}}] \gamma_{j'}=\beta_{00}(\mathbf{r}) \gamma_{j'}$. The effect of local optical pumping on spin squeezing will be described in the following sections.

\subsection{Multilevel structure of Cesium atoms in the ensemble}

For weak coupling, the many-body state of the large ensemble is well described in the Gaussian approximation, fully determined by one and two-body correlation.  The dynamics of these correlations are governed by the adjoint form of the SME, Eq. (\ref{SME}), and form a closed set of equations that can be integrated together. In Ref. \cite{Baragiola2014} the constituent atom were spin-$\tfrac{1}{2}$ so the collective spin operators themselves were used; here, we take into account the multilevel nature of each Cs atom. Below we present the formalism for the symmetric collective situation and describe the projection onto spatial modes for the full three-dimensional model. 

We treat the internal state space of each individual Cs atom as a 3-level system (qutrit) with basis states, 
\begin{equation}
\begin{aligned}
| \hspace{-1mm}\uparrow\rangle \coloneqq |6S_{1/2}, f=4, m_{x}=4 \rangle \hspace{1mm}\\ 
| \hspace{-1mm}\downarrow\rangle \coloneqq |6S_{1/2}, f=4, m_{x}=3 \rangle  \hspace{1mm} \\
|T \rangle \coloneqq |6S_{1/2},  f=4, m_{x}=2 \rangle .
\end{aligned}
\end{equation}
This basis is chosen as follows. Initially, each atom is optically pumped into ``fiducial" the internal state, $\ket{\uparrow}$. The QND measurement, following the Faraday interaction, is dominated by symmetrically entangling atoms in $\ket{\uparrow}$ with atoms in the ``coupled state," $\ket{\downarrow}$.  We include the ``transfer state," $\ket{T}$, to account for transfers of coherence that occur due to partial indistinguishability of scattered photons during optical pumping~\cite{Norris2012}. In principle one could continue this process and construct an complete ``Faraday basis" for the internal state of a single Cs atom. For weak coupling, three states suffice. Using this truncated Hilbert space, atoms optically pumped to the $f=3$ manifold are lost as are atoms that exit the qutrit subspace within the $f=4$ manifold. These effects are accounted for by projecting the Faraday interaction and optical pumping map into the qutrit basis.

\subsection{Collective operators}

In order to calculate the spin squeezing parameter and generate simulated experimental polarimeter signals, Eq. (\ref{eq:polarimetersignal}), we require the collective spin moments $\langle{ \hat{F}_x \rangle}$,
$\langle{ \hat{F}_z \rangle}$ and $\Delta \hat{F}_z^2 $. These moments are decomposed in terms of collective population and coherence operators, defined over the qutrit subspaces in the atoms,
	\begin{align}\label{Quadratures}
	\hat{N}_{{i}} \coloneqq  \sum_{n=1}^{N_A} \hat{n}_i^{(n)} \, , \quad  
	\hat{X}_{i j} \coloneqq  \sum_{n=1}^{N_A}\hat{x}_{ij}^{(n)} \, , 
	\end{align}
where the single-atom operators are
	\begin{subequations} \label{singleatom}
	\begin{align}
		\hat{n}_i \coloneqq & | {i} \rangle \langle {i} | \, , \\
		\hat{x}_{i j} \coloneqq & \tfrac{1}{\sqrt{2}} \big( | {i} \rangle \langle {j} | + | {j} \rangle \langle {i} | \big) \, , 
	\end{align}
	\end{subequations}
with $\{ {i}, {j} \} \in \{ \uparrow, \downarrow, {T} \}$ and $j > i$ in $\hat{x}_{ij}$ to avoid redundancy. The collective spin operators $\hat{F}_{x}$ and $\hat{F}_{z}$ that appear in Eq. (\ref{eq:squeezingparam}) relate to these operators in the following way \cite{NorrisThesis2014}, 
\begin{subequations} \label{eq:spinoperators}
\begin{align}
	\hat{F}_{x} \approx & f  \hat{N}_{\uparrow}  + (f-1)   \hat{N}_{\downarrow}  + (f-2)  \hat{N}_{T}   \, ,\\
	\hat{F}_z \approx & v_\uparrow  \hat{X}_{\uparrow \downarrow} + w_\uparrow  \hat{X}_{ \downarrow T} \, ,
\end{align}
\end{subequations}
where the coefficients $v_\uparrow \coloneqq \sqrt{ (\Delta f_z^2)_\uparrow }$ and $w_\uparrow \coloneqq \sqrt{ 2 (\Delta f_z^2)_\downarrow - 2 (\Delta f_z^2)_\uparrow  }$ depend on single-atom variances of $\hat{f}_{z}$ under $| \hspace{-1mm}\uparrow\rangle$ and $| \hspace{-1mm}\downarrow\rangle$ \cite{Norris_thesis}. 
This gives for the variance,
	\begin{align}
\Delta F_{z}^2 
 & \approx v_\uparrow^2 \Delta X_{\uparrow \downarrow }^2 +2 v_\uparrow w_\uparrow \langle \Delta \hat{X}_{ \uparrow \downarrow } \Delta \hat{X}_{\downarrow  T  } \rangle + w_\uparrow^2 \Delta X_{\downarrow  T }^2 ,
	\end{align}
where $\Delta \hat{A} \coloneqq \hat{A} - \langle \hat{A} \rangle$.
Equations (\ref{eq:spinoperators}) are appropriate for short times (several photon scattering times) when the majority of the population resides in the fiducial state and little is lost outside the qutrit subspace.

\subsection{Equations of motion for the spatially inhomogeneous collective operators}
We  combine the three-dimensional model with the multilevel description of the spin-4 Cs atoms and present the coupled set of equations for the collective operators. As discussed in detail in Refs.~\cite{Baragiola2014, BaragiolaThesis2014, NorrisThesis2014}, in order to account for the local optical pumping, we divide $z$-direction (along the probe's propagation) into coarse-grained longitudinal slices of width $\Delta z$. Then, within each longitudinal slice the collective operators are projected into the set of transverse LG modes. This approximate longitudinal-mode decomposition improves as $\Delta z$ decreases. 

A single-body collective operator labeled by the transverse $pl$-mode decomposes longitudinally as
	\begin{align}
		\hat{O}^{pl} = \sum_{n=1}^{N_A} \beta_{pl}(\mathbf{r}_n) \hat{o}^{(n)} = \sum_k \hat{O}^{pl}(z_k),
	\end{align}
where $\hat{O}^{pl}(z_k) = \sum_{n_k} \beta_{pl}(\mathbf{r}_{n_k}) \hat{o}^{(n_k)}$, with this sum running only over those atoms in the longitudinal slice centered at $z_k$ with width $\Delta z$. 
The collective, fundamental-mode spin operators that contribute to the polarimeter signal, Eq. (\ref{eq:polarimetersignal}), and spin-squeezing parameter, Eq. (\ref{eq:squeezingparam}), are single-body and decompose spatially as
	\begin{subequations} \label{eq:spinwavedecomposition}
	\begin{align} 
	\hat{F}^{00}_{x} \approx & \sum_{k} \Big[ f  \hat{N}^{00}_{\uparrow}(z_k)  + (f-1) \hat{N}^{00}_{\downarrow}(z_k)  \\
	& + (f-2)  \hat{N}^{00}_{T}(z_k) \Big]  \, , \nonumber \\
	\hat{F}^{00}_z \approx &  \sum_k \Big[ v_\uparrow  \hat{X}^{00}_{\uparrow \downarrow}(z_k) + w_\uparrow  \hat{X}^{00}_{ \downarrow T}(z_k) \Big] \, . 
	\end{align}
	\end{subequations}
A collective operator such as $(\hat{F}^{00}_z)^2$ involves two sums over longitudinal modes, as it describes correlations between atoms both within a single longitudinal slice and also between different slices. Thus it contains both one-body and two-body terms, each of which is affected differently by local optical pumping.
	
Local optical pumping, Eq. (\ref{eq:opmap}), generates dynamics for a single-body operators,
	\begin{align}
		\frac{d}{dt} &  \hat{o}^{(n)} \Big|_{\rm op }= \sum_{j'} \gamma_{j'}(\mathbf{r}_n) \mathcal{D}^{(n) \dagger}_{j'}[\hat{o}^{(n)}], 
	\end{align}	
where the dagger on the optical pumping map indicates that it is in adjoint form (appropriate for Heisenberg-picture evolution of operators). 
For a two-body operator optical pumping leads to a decay of correlations driven by the dynamics,
\begin{align} \label{eq:twobodydecay}
	\frac{d}{dt} \hat{o}^{(m)} \hat{v}^{(n)} \Big|_{\rm op} = &\sum_{j'} \bigg\{ \gamma_{j'}(\mathbf{r}_m)\mathcal{D}_{j'}^{(m)\dagger}[\hat{o}^{(m)}] \hat{v}^{(n)} \\
	& + \gamma_{j'}(\mathbf{r}_n) \hat{o}^{(m)} \mathcal{D}_{j'}^{(n)\dagger}[ \hat{v}^{(n)}] \bigg\} \nonumber
\end{align}
The optical pumping map for each local operator is weighted by the local scattered rate of the associated atom.
The fact that the optical pumping acts locally breaks the collective symmetry of the spin waves in the SME, Eq. (\ref{SME}), and leads to a coupling of collective spin waves at the same longitudinal slice but in different LG modes. Here we use the optical pumping map after the RWA with respect to the bias magnetic field has been applied, as described at the end of Sec. IIA. 

As discussed in detail in Ref.~\cite{Baragiola2014} the evolution of single-body and two-body collective operators couple in a complicated way between spatial modes. Nevertheless, by projecting the effects of the optical pumping map into the qutrit basis we find the following closed, deterministic set of equations in Gaussian approximation \cite{NorrisThesis2014},
\begin{widetext}
\begin{subequations} \label{eq:EOMs}
\begin{align}
	\frac{d}{dt} & \langle \hat{N}_{i}^{pl} (z_k) \rangle = c_{p'l'}^{pl}(z_k) \sum_{j'} \sum_\ell \sum_{p',l'} \gamma_{j'} \Tr[ \mathcal{D}_{j'}^\dagger[\hat{n}_i] \hat{n}_\ell ] \big\langle \hat{N}_\ell^{p'l'}(z_k) \big\rangle \, ,\\
	\frac{d}{dt} & \langle \hat{X}_{ij}^{pl} (z_k) \hat{X}_{i'j'}^{p'l'} (z_{k'}) \rangle_{\rm s} =  \\
	& - \kappa \sum_{k,k'} \Big[ v_\uparrow \big\langle \hat{X}_{\uparrow \downarrow}^{00} (z_k'') \hat{X}_{ij}^{pl} (z_{k'''}) \big\rangle_{\rm s} + w_\uparrow \big\langle \hat{X}_{ \downarrow T}^{00} (z_k'') \hat{X}_{ij}^{pl} (z_{k}) \big\rangle_{\rm s} \Big] 
	 \Big[ v_\uparrow \big\langle \hat{X}_{\uparrow \downarrow}^{00} (z_k''') \hat{X}_{i'j'}^{p'l'} (z_{k'}) \big\rangle_{\rm s} + w_\uparrow \big\langle \hat{X}_{ \downarrow T}^{00} (z_k''') \hat{X}_{i'j'}^{p'l'} (z_{k'}) \big\rangle_{\rm s} \Big] \nonumber \\
	& +  \sum_{j'} \sum_{\ell, m} \sum_{p'',l''} \gamma_{j'} \left( c_{p''l''}^{pl}(z_k) \Tr [\mathcal{D}^\dagger_{j'}[\hat{x}_{ij}] \hat{x}_{\ell m}] \big\langle \hat{X}_{i'j'}^{p'l'} (z_{k'}) \hat{X}_{\ell m}^{p''l''} (z_{k}) \big\rangle_{\rm s}  + c_{p''l''}^{p'l'}(z_{k'}) \Tr [\mathcal{D}^\dagger_{j'}[\hat{x}_{i'j'}] \hat{x}_{\ell m}] \big\langle \hat{X}_{ij}^{pl} (z_k'') \hat{X}_{\ell m}^{p''l''} (z_{k'}) \big\rangle_{\rm s} \right) \nonumber \\
	& +  \, \delta_{k,k'}  \sum_{j'} \sum_{\ell}  \sum_{p'',l''} \gamma_{j'} g_{p''l''}^{pl p'l'}(z_{k'}) \Tr [\mathcal{N}_{j'}[\hat{x}_{i j}, \hat{x}_{i' j'}] \hat{n}_{\ell} ] \big\langle \hat{N}_{\ell m}^{p''l''} (z_{k'}) \big\rangle, \nonumber 
\end{align}
\end{subequations}
where $\{ \ell,m\} \in \{ \uparrow, \downarrow, T \}$, we use symmetrized moments, $\langle \hat{A} \hat{B} \rangle_{\rm s} \coloneqq \frac{1}{2}\langle \hat{A} \hat{B} + \hat{B} \hat{A} \rangle$. The superoperator in the final line, which arises from the two-body decay map in Eq. (\ref{eq:twobodydecay}), is
	\begin{align}
		\mathcal{N}_{j'}[\hat{a}, \hat{b}] \coloneqq \frac{1}{2} \left( \mathcal{D}^\dagger_{j'}[\{\hat{a},\hat{b} \}_+] - \{ \mathcal{D}^\dagger_{j'}[\hat{a}],\hat{b} \}_+ - \{ \hat{a},\mathcal{D}^\dagger_{j'}[\hat{b}] \}_+  \right) \, ,
	\end{align}
where $\{ \hat{a},\hat{b} \}_+ =  \hat{a}\hat{b} +  \hat{b}\hat{a}$ is the anticommutator. 
The coefficients that describe the projection of the optical pumping into the LG modes are \cite{Baragiola2014, BaragiolaThesis2014, NorrisThesis2014},
	\begin{align}
		c_{p'l'}^{pl}(z) \coloneqq & \frac{1}{A} \int d^2 \mathbf{r}_\perp [u_{00}(\mathbf{r}_\perp, z)]^2 u^*_{pl}(\mathbf{r}_\perp, z) u^*_{p'l'}(\mathbf{r}_\perp, z) \, , \\
		g_{p''l''}^{pl p'l'}(z) \coloneqq & \frac{1}{A} \int d^2 \mathbf{r}_\perp u_{00}(\mathbf{r}_\perp, z) u_{p''l''}(\mathbf{r}_\perp, z) \beta_{pl}(\mathbf{r}_\perp, z) \beta_{p'l'}(\mathbf{r}_\perp, z).
	\end{align}

The single-body quadratures that contribute to the spin waves, $\langle \hat{F}_z^{pl}\rangle$, including the squeezed spin wave, $\langle \hat{F}_z^{00}\rangle$, depend on the measurement record and evolve according the adjoint-form SME,
	\begin{align} \label{eq:XEOM}
		d \langle \hat{X}_{ij}^{pl} (z_k) \rangle =&  \sqrt{\kappa} \langle \Delta \hat{X}_{ij}(z_k) \Delta \hat{F}^{00}_{z} \rangle_{\rm s} dW \\
		& +  \sum_{j'}\sum_{\ell}\sum_{p',l'}  \gamma_{j'} c_{p'l'}^{pl}(z_k) \bigg\{   \Tr [\mathcal{D}_{j'}[\hat{x}_{ij}] \hat{n}_{\ell}] \langle \hat{N}_{\ell}(z_k) \rangle + \sum_{m} \Tr [\mathcal{D}_{j'}[\hat{x}_{ij}] \hat{x}_{\ell m}] \langle \hat{X}_{\ell m}(z_k) \rangle \nonumber \bigg\} dt \nonumber .
	\end{align}
The full set of operator equations of motion is found by expanding $\Delta \hat{F}^{00}_{z}$ in terms of the collective quadratures using Eq. (\ref{eq:spinwavedecomposition}).

Finally, every atom in the cloud is optically pumped to the fiducial state $\ket{\uparrow}$, giving initial conditions for the above equations,
	\begin{align}
	\langle \hat{N}_{i}^{pl} (z_k) \rangle(t_0) = & \int^{z_k + \frac{\Delta z}{2}}_{z_k - \frac{\Delta z}{2}} dz \int d^2 \mathbf{r}_\perp \eta(\mathbf{r}_\perp, z) \beta_{pl}(\mathbf{r}_\perp, z) \bra{\uparrow} \hat{n}_i \ket{\uparrow} \, , \\
	\langle \hat{X}_{ij}^{pl} (z_k) \rangle(t_0) = &  \int^{z_k + \frac{\Delta z}{2}}_{z_k - \frac{\Delta z}{2}} dz \int d^2 \mathbf{r}_\perp \eta(\mathbf{r}_\perp, z) \beta_{pl}(\mathbf{r}_\perp, z) \bra{\uparrow} \hat{x}_{ij} \ket{\uparrow} \, , \\
	\langle \hat{X}_{ij}^{pl} (z_k) \hat{X}_{i'j'}^{p'l'} (z_{k'}) \rangle_{\rm s}(t_0) = & \delta_{k,k'}  \int^{z_k + \frac{\Delta z}{2}}_{z_k - \frac{\Delta z}{2}} dz \int d^2 \mathbf{r}_\perp \eta(\mathbf{r}_\perp, z) \beta_{pl}(\mathbf{r}_\perp, z) \bra{\uparrow} \hat{x}_{ij} \hat{x}_{i'j'}\ket{\uparrow}_{\rm s}.
	\end{align}

Into this model we input the experimental parameters: (i) probe powers, (ii) probe detunings, (iii) probe spatial modes, (iv) atomic-cloud peak density, and (v) atomic-cloud-density $e^{-2}$ variances. 
We numerically integrate Eqs. (\ref{eq:EOMs}) and Eq. (\ref{eq:XEOM}), whose solutions are used to reconstruct the first- and second-order collective spin moments, including the squeezed spin-wave variance $(\Delta F_z^{00})^2$, using Eqs. (\ref{eq:spinwavedecomposition}). 

With the equations of motion for these one and two-body correlations, we can calculate the metrological squeezing parameter,
\begin{align} \label{eq:squeezingparam}
	\xi_m^2 = \left( \frac{\Delta \phi }{ \Delta \phi_{CS} } \right)^2 = 2f \frac{ N_1^2 }{N_2} \frac{( \Delta F_z^{00})^2 }{ \langle \hat{F}_x^{00} \rangle^2},
\end{align}
where $\langle \hat{F}_x^{00} \rangle_{CS} = N_1 f$ and $( \Delta F_z^{00})^2 = N_2f/2$, 
and the effective atom numbers are $ N_K \coloneqq \int d^3 \mathbf{r} \, \eta(\mathbf{r}) |u_{00}|^{2K}$ \cite{Baragiola2014}.    This gives a theoretical prediction of the squeezing as a function of time, as shown in Fig. 3b (blue curve).

We can also directly numerically simulate the measurement record using Eq. (\ref{eq:polarimetersignal}).  A simulation of the  stochastically varying spin moment $\langle \hat{F}_z^{00} \rangle$  also allows us to directly generate simulated data for each integration of the equations. We analyze a collection of simulated measurement records in the same way we do bonafide measurement records from the experiment as shown in Fig. 3b (red curve).

\end{widetext}

\end{document}